# Calibration of DOI-Capable PET Detector Panels Using Uncollimated Front-Face Irradiation

Francisco E. Enríquez-Mier-y-Terán[a,b,*], Andre Z. Kyme[b,c], Steven R. Meikle[a,b,d]

[a]School of Health Sciences, Faculty of Medicine and Health, The University of Sydney, Sydney, NSW 2050, Australia

[b]Brain and Mind Centre, The University of S ydney, Sydney, NSW 2050, Australia

[c]School of Biomedical Engineering, Faculty of Engineering, The University of Sydney, Sydney, NSW 2006, Australia

[d]Sydney Imaging Core Research Facility, The University of Sydney, Sydney, NSW 2008, Australia

*Corresponding author.
E-mail address: francisco.enriquez@sydney.edu.au (F.E. Enríquez-Mier-y-Terán)

## Abstract

*Objective*. Gold-standard depth-of-interaction (DOI) calibration using collimated gamma-ray irradiation is time-consuming and requires complex experimental setups, making it impractical for system-level calibration of detector arrays. This work investigates an efficient method for DOI and energy calibration of detector panels using uncollimated irradiation, in which gamma rays are incident parallel, or nearly parallel, to the crystal depth direction.

*Approach.* The 511-keV photopeak location on a dual-ended readout PET detector block as a function of crystal depth was first evaluated using both collimated and uncollimated $^{22}$Na source irradiation to determine whether the latter yields comparable estimates when DOI calibration parameters are known. A $4 \times 4$ dual-ended readout PET detector panel was then assembled, and three detector blocks were calibrated using the gold-standard method. Two DOI calibration approaches based on uncollimated irradiation—a physics-informed model and a multilayer perceptron (MLP)—were compared against the gold-standard calibration. Finally, the detector panel was calibrated for DOI and energy using the MLP-based approach.

*Main Results*. The median relative root mean squared error (RMSE) between second-order polynomial fits to the depth-dependent photopeak location obtained using collimated and uncollimated irradiation was 1%, demonstrating that uncollimated irradiation provides reliable estimates when accurate DOI calibration parameters are available. The RMSE between the gold-standard DOI calibration and the physics-informed and MLP-based approaches ranged from 0.38–0.58 mm and 0.36–0.61 mm, respectively, with no statistically significant difference. However, DOI resolution estimates from the MLP-based approach outperformed those of the physics-informed model, favouring its use for full detector panel calibration. After saturation correction, the detector panel achieved a mean energy resolution of 15.6% and a DOI resolution of 2.0 mm.

*Significance*. The proposed MLP-based calibration requires only a single uniform 511-keV source irradiation, making it simple to implement and sufficiently robust for in situ calibration of DOI-capable PET detector arrays.

## Introduction

Small-animal and organ-dedicated positron emission tomography (PET) imaging requires sub-millimetre spatial resolution to enable accurate quantification of molecular processes occurring within small anatomical structures. Accordingly, these PET systems typically use highly pixelated or (semi-)monolithic scintillation detectors arranged in narrow-bore geometries. Such geometries require accurate 3D gamma-ray interaction positioning to mitigate parallax errors, which in turn requires time-consuming detector calibration procedures (Enríquez-Mier-y-Terán *et al* 2025).

Most methods to estimate the gamma-ray depth of interaction (DOI) in a scintillation crystal require dedicated DOI calibration (Ito *et al* 2011, Pizzichemi *et al* 2016). Typically, a DOI calibration requires lateral access to the crystal array and linear actuators to precisely position either a slab-based detector (Fig. 1a) or another physical collimator, such as tungsten blocks, to confine the gamma-ray beam to a narrow region of the array. Measurements are usually acquired at two or more irradiation depths with long acquisition times at each depth to ensure sufficient counting statistics in all crystals. Although this approach is considered the gold standard for DOI calibration, it is impractical for calibrating the many detector blocks that constitute a typical PET system. Moreover, as PET systems evolve towards both higher resolution and higher sensitivity, designs featuring reduced detector pitch and increased axial coverage may become increasingly attractive (Vandenberghe *et al* 2020, Sanaat *et al* 2024, Allen *et al* 2024), and for such configurations side irradiation is highly impractical if not impossible, due to lack of access to detector elements near the centre of a many-element array.

Detector DOI calibration is required not only for accurate 3D gamma-ray positioning but also for other depth-dependent detector calibrations. For example, depending on the inter-crystal reflector material, the position of the gamma-ray photopeak in the energy spectrum may vary as a function of crystal depth (Borghi *et al* 2016, Kuang *et al* 2019, Cucarella *et al* 2025, Enríquez-Mier-y-Terán *et al* 2025). If uncorrected, this effect leads to an apparent broadening of the photopeak, thereby degrading the detector energy resolution and PET system performance. In time-of-flight (TOF) PET, DOI information is also used to correct timestamp shifts arising from differences in the path length travelled by optical photons which would otherwise degrade the system's coincidence time resolution (CTR) (Yoshida *et al* 2023, Nadig *et al* 2024, Terragni *et al* 2025, Kratochwil *et al* 2025)

The efficiency and feasibility challenges of performing system-level DOI calibration have motivated methods that do not rely on lateral access to the crystal array and/or complex irradiation setups. For dual-ended readout detectors, Shao *et al* (2008) proposed a DOI calibration method based on uncollimated (uniform) lateral irradiation of the crystal array and the cumulative distribution function of the DOI ratio, defined as ratio of the event energy measured by photodetector 2 ($E_2$) to the total event energy measured by photodetectors 1 and 2 ($E_1$+$E_2$):

$$DOI_{ratio} = \frac{E_2}{E_1 + E_2} \tag{1}$$

Since this approach still required lateral access to the crystal array, the methodology was extended to exploit the intrinsic background radiation of LSO/LYSO crystals instead of external sources (Bircher and Shao 2012) and, subsequently, to uncollimated front-face gamma-ray irradiation (Fig. 1b) (Stringhini *et al* 2016). Using a dual-ended readout PET detector and uncollimated irradiation, Yang *et al* (2008) proposed estimating the crystal ends by identifying the midpoint of the rising and falling edges of the DOI ratio histogram. This approach assumes a linear relationship between DOI ratio and depth which may not always hold.

For monolithic detectors, Kuhl *et al* (2023) reported using front-face oblique collimated irradiation together with in-plane $(x, y)$ positioning information to estimate DOI. However, this approach is prone to error resulting from mechanical tolerances and uncertainties in $(x, y)$ positioning. Lerche *et al* (2008) proposed modelling the DOI histogram—obtained from front-face irradiation of the crystal—as the convolution of gamma-ray attenuation with a constant DOI resolution. This physics-informed model enables retrieval of both the crystal ends and the DOI resolution (in DOI space), which can subsequently be used to estimate the detector DOI resolution in millimetres. Unlike previous approaches, this method provides a direct measurement of DOI resolution. This is important in PET system development, as it not only enables depth-dependent detector calibration but also informs DOI binning strategies during image reconstruction and facilitates more consistent comparisons between high-resolution PET systems. In this work, however, we show that despite these advantages, DOI resolution can be estimated more accurately using a simple machine-learning approach.

Since 2020, a variety of machine-learning methods that predict DOI on an event-by-event basis have been proposed (Zatcepin *et al* 2020, LaBella *et al* 2021, Freire *et al* 2022, Kuhl *et al* 2023, Dai *et al* 2025, Kinyanjui *et al* 2025). While these methods avoid assuming a predefined relationship between the detector metric and DOI, they typically require large training datasets to adequately characterise the depth-dependent detector response, as the DOI relationship must be learned directly from the data rather than being constrained by a low-dimensional parametric model, potentially restricting their adoption for system-level calibration. Consequently, there is a need for machine-learning approaches that combine the flexibility of data-driven methods with the simplicity and scalability required for practical system-level DOI calibration.

The goal of this work was to develop and validate a practical and robust approach for system-level DOI calibration to enable accurate gamma-ray positioning and other depth-dependent detector corrections. Specifically, our objectives were: (i) to evaluate whether uncollimated front-face detector irradiation yields photopeak location estimates comparable to those obtained using the gold-standard side-irradiation approach, using a previously characterised dual-ended readout PET detector (Enríquez-Mier-y-Terán *et al* 2025); (ii) to compare the performance of two methods for DOI calibration and DOI resolution estimation: the physics-informed DOI model proposed by Lerche *et al* (2008) and a multilayer perceptron (MLP)-based approach; and (iii) to demonstrate the practical value of the MLP-based approach for DOI calibration of an entire detector panel intended for sub-millimetre PET imaging applications.

## 2. Methods

### 2.1. Single Detector Irradiation

A dual-ended readout PET detector was assembled to compare the photopeak location and detector energy resolution as a function of the crystal depth using traditional electronically collimated (slab-based) irradiation and uncollimated front-face irradiation (Fig. 1). The test detector consisted of an array of 23 × 23 LYSO crystals (Boya Advanced Materials Company Ltd., China) of dimension 0.785 mm × 0.785 mm × 20 mm (pitch = 0.85 mm), optically isolated using Lumirror Toray reflector. At each end of the crystal array, a 6 × 6 MicroFJ-30035 silicon photomultiplier (SiPM) array (ONSEMI, USA) was optically coupled via 1.2 mm thick optical-grade PMMA sheets (Eljen Technology, USA) and EJ-550 optical grease (Eljen Technology, USA).

The 72 SiPM signals are reduced to 8 signals (an $X^+$, $X^-$, $Y^+$ and $Y^-$ Anger logic input signal for each SiPM array) prior to being fed into different channels of a TOFPET2 ASIC (PETSys Electronics S.A., Portugal) for data digitisation. The detector bias voltage was set to 27.5 V and the ASIC threshold parameters, *vth_t1*, *vth_t2* and *vth_e*, set to 50, 20, and 15 DAC units, respectively. Only events triggering the eight ASIC channels and within a 30 ns time window from the first triggered channel were considered valid. The gamma ray $(x, y)$ positioning was estimated using Anger logic before being assigned to its nearest crystal centre using Voronoi diagrams. The DOI was subsequently estimated from the DOI ratio, where gamma ray energy is measured as the sum of the eight recorded signals. Further details can be found in (Enríquez-Mier-y-Terán *et al* 2025). In all cases, experiments were conducted in a light-tight environment at a controlled temperature of $22.0 \pm 0.2$ ℃.

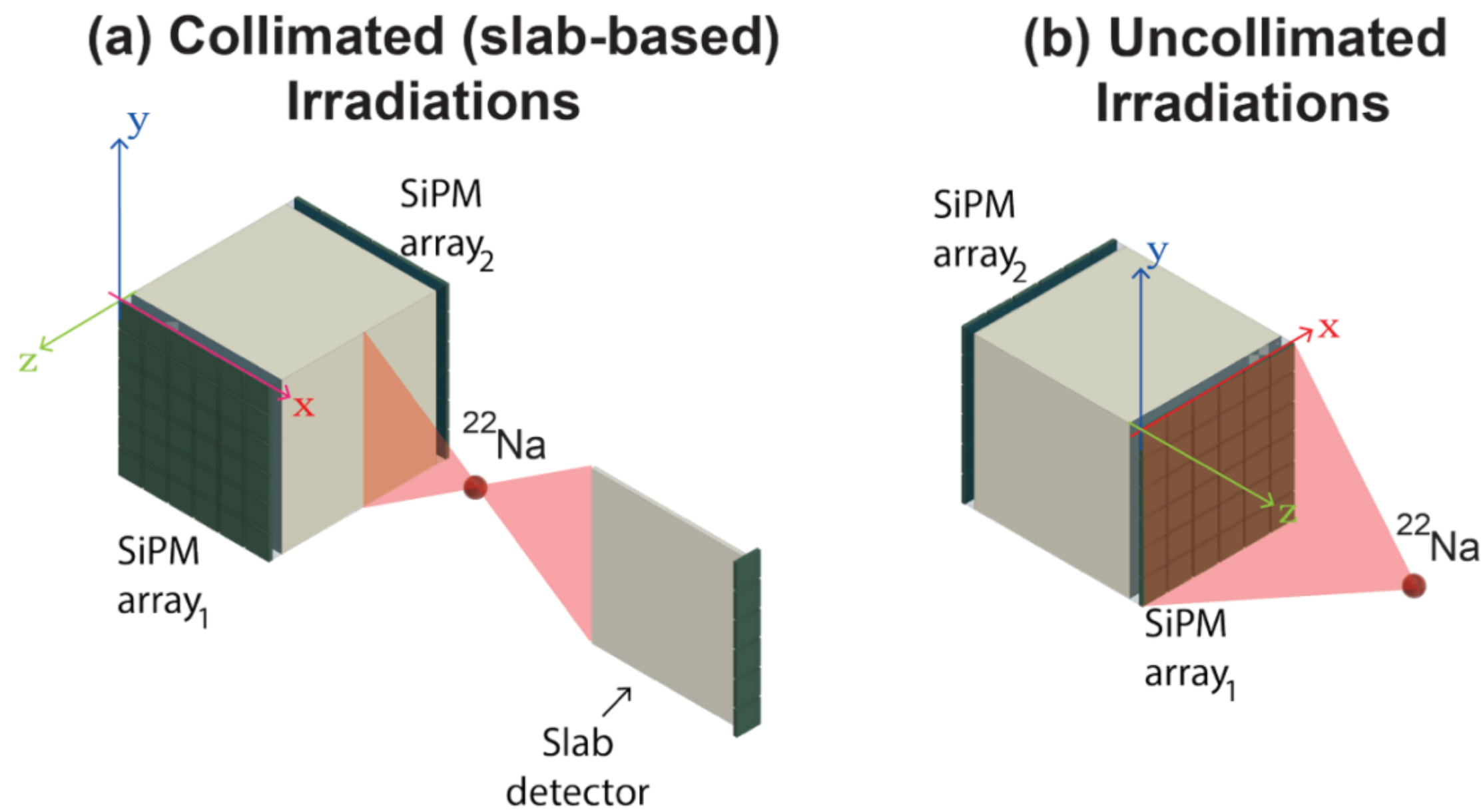

Fig 1. Experimental setups for the single-detector irradiation. (a) Gold-standard irradiation using a slab detector for electronic collimation. (b) Front-face irradiation using an uncollimated source.

### 2.1.1. Electronically Collimated Irradiation

A Teflon-wrapped LYSO slab (0.5 mm × 20 mm × 20 mm), coupled to a 1 × 6 MicroFJ-30035 SiPM array, was assembled to electronically collimate a $^{22}Na$ point source (Fig. 1a). Irradiation was performed at five depths (2, 6, 10, 14, and 18 mm), each with a duration of 3 hours. At each depth, crystal-wise photopeak locations (in ASIC units) and energy resolution values (without saturation correction) were estimated by fitting a Gaussian function to the 511-keV peak and extracting its mean and full-width at half maximum (FWHM), respectively.

DOI ratios were calculated on an event-by-event basis. These ratios were subsequently histogrammed based on crystal and depth. Gaussian functions were fitted to the resulting DOI ratio histograms and the extracted mean values fitted using a linear function to extract the slope ($k^{truth}$) and intercept ($i^{truth}$), thereby defining the DOI calibration. For each crystal, the DOI resolution in millimetres ($DOI_{mm}^{truth}$) was obtained by averaging the FWHM across all depths and converting it to distance units using the calibration parameters.

### 2.1.2. Uncollimated Irradiation

The $^{22}Na$ source was centred with respect to the detector front face—the SiPM array$_1$ $xy$ plane—and positioned 40 mm from it in the $z$ direction (Fig. 1b). The test detector was irradiated for 30 minutes followed by a 30-minute background radiation acquisition (i.e., without the presence of the $^{22}Na$ source). The DOI calibration parameters obtained in Section 2.1.1 were used to virtually segment each crystal in the array along the $z$ (depth) dimension into eight virtual crystals of 2.5 mm thickness. Energy spectra were constructed for the virtual crystals by subtracting the background energy spectrum from the $^{22}$Na energy spectrum. Gaussian distributions were fitted to the 511-keV peaks to estimate photopeak location and energy resolution.

## 2.2. Detector Panel Assembly

The PET detector panel comprised 4 × 4 test detector blocks (Fig. 2a,b). This panel forms part of the motion-adaptive MousePET system under development in our group (Kyme *et al* 2017, Enríquez-Mier-y-Terán *et al* 2021). Since mechanical stability of the detector panel is vital for this motion-adaptive system, the 4 × 4 crystal arrays (separated by 0.9 mm Al shims) were laterally bonded together and to two aluminium blocks using DP420 structural adhesive (3M™ Scotch-Weld™, USA) (Fig. 2b). Furthermore, a single 1.2 mm thick PMMA sheet as a light guide was glued to each end of the 4 × 4 crystal array using EPO-TEK 301 optical adhesive (Epoxy Technology Inc., USA) (Fig. 2b,c). To reduce optical crosstalk between neighbouring detectors, 0.8 mm wide × 0.5 mm deep grooves were machined into the PMMA and filled with highly reflective white paint (Eljen Technology, USA) (Fig. 2c). EJ-550 optical grease (Eljen Technology, USA) was used as the optical coupling between the SiPM arrays and the extended PMMA sheets.

The 72:8 signal multiplexing scheme (Section 2.1) was applied to every detector. Per panel, 128 signals are digitised using two TOFPET2 ASICs. The 32 SiPM arrays in the panel are biased using a single power line, with every array biased at 27 V. This bias voltage produced visibly better flood maps compared to 27.5 V due to detector gain variations across the panel. ASIC thresholds *vth_t1*, *vth_t2* and *vth_e* were fixed at 50, 20, and 20 DAC units, respectively, for all detectors. As in Section 2.1, valid detector events were only those that triggered the eight signal channels within a time window of 100 ns. A temperature monitoring system based on thermistors was developed to monitor the temperature of each SiPM array in the detector panel. The SiPM arrays were kept at a temperature of 22.0 ± 0.5 ℃ during the experiments.

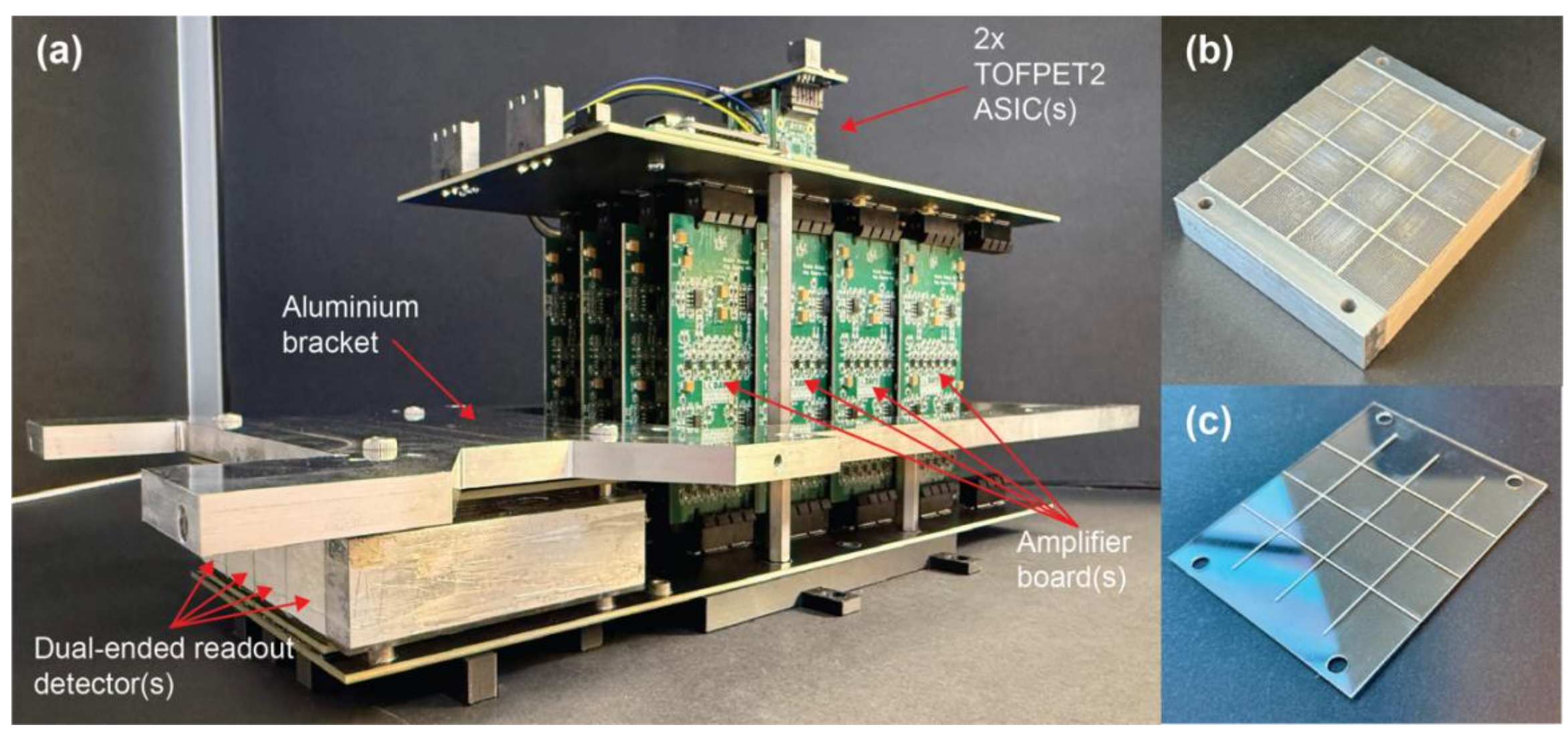

Fig. 2. (a) PET detector panel: (left) the 4 x 4 detector array, including the SiPM array boards, PMMA sheets and LYSO crystals; (right) signal multiplexing and amplification circuit boards interfacing the detectors to two TOFPET2 ASICs. (b) 4 x 4 crystal array and PMMA sheet after these were structurally and optically bonded. (c) Slitted PMMA sheet with reflective white paint.

## 2.3. Detector Panel Irradiation

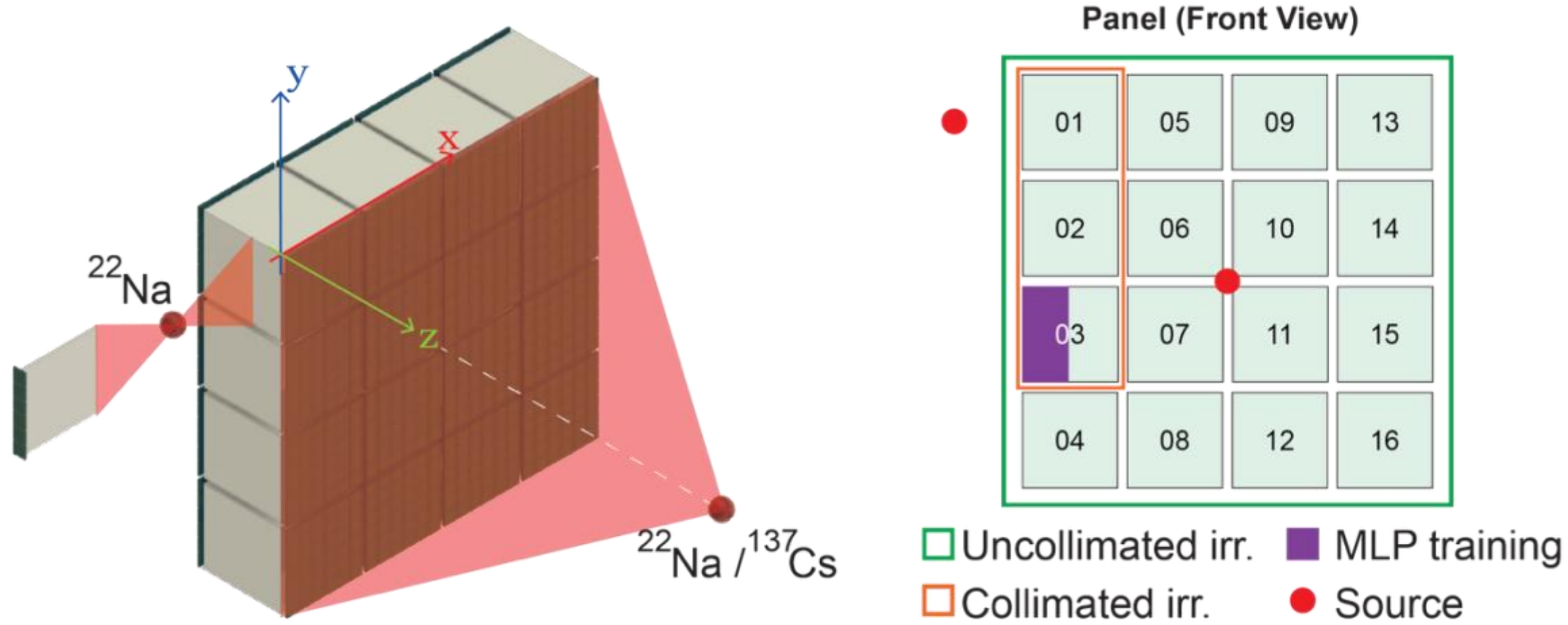

Fig 3. Detector panel irradiation setups. Detectors are numbered from 01 to 16 for clarity. Detector panel calibration (see Section 2.3.2) was performed using uncollimated irradiation from a $^{22}$Na and $^{137}$Cs source, respectively, centred with respect to the detector panel front face and positioned 60 mm from it in the $z$ direction. The accuracy of the proposed methods for DOI calibration parameters and resolution (see Section 2.3.1) was evaluated against gold-standard collimated irradiation performed on three detectors from the panel (orange frame). To increase the training data for the MLP method (see Section 2.3.1), DOI ratio histograms (from the uncollimated $^{22}$Na source irradiation) and ground-truth DOI calibration and resolution parameters (from collimated irradiation) from 230 crystals from one detector (in purple) were used.

### 2.3.1. DOI Calibration and Resolution

Because of the detector panel configuration, a side-wise collimated irradiation (Section 2.1.1) is unlikely to penetrate the central $2 \times 2$ detector grid (Fig. 3). Accordingly, two approaches necessitating only uncollimated irradiation (Section 2.1.2) were investigated for DOI calibration and resolution estimation in each detector: (i) a physics-informed DOI model (Lerche *et al* 2008) and (ii) our proposed multilayer perceptron (MLP) model. To evaluate the accuracy of these methods, three detectors in the panel were irradiated at three different depths (2, 10 and 18 mm) using side-wise collimated irradiation (Fig. 3; orange frame) and the calibration and resolution results compared.

2.3.1.1 Physics-informed model. In the physics-informed DOI model (Lerche *et al* 2008), the DOI ratio histogram ($D$) is modelled as the convolution of the detector intrinsic DOI resolution, assumed to be depth-independent and Gaussian-distributed, and the gamma-ray attenuation as a function of crystal depth (Fig. 4a):

$$D(d) = A\, e^{-\frac{\mu^*_{LYSO}(d-a)}{b-a}} \left[ \mathrm{erf}\left(\frac{b-d}{\sqrt{2}\sigma_m}\right) - \mathrm{erf}\left(\frac{a-d}{\sqrt{2}\sigma_m}\right) \right] \tag{2}$$

where $A$ is a scaling factor, $d$ denotes the interaction depth in DOI space, and $\sigma_m$ is the standard deviation of the Gaussian DOI resolution. The parameter $\mu^*_{LYSO}$ represents the attenuation coefficient for 511-keV gamma rays in LYSO ($\mu_{LYSO}$) scaled by the crystal thickness (i.e., $\mu^*_{LYSO} = 20\ mm \times \mu_{LYSO}$), and $a$ and $b$ are the proximal (0 mm) and distal (20 mm) crystal ends in the DOI space, respectively.

For each crystal, the DOI calibration parameters—slope ($k^{model}$) and intercept ($i^{model}$) —were obtained from a linear fit between the crystal end positions in physical space ($[0,20]$ mm) and their corresponding locations in DOI space $[a, b]$. Following (Lerche *et al* 2008), the DOI resolution in millimetres ($DOI^{model}_{mm}$) was obtained by scaling $\sigma_m$:

$$DOI^{model}_{mm} = 2.35\sigma_m \frac{20\ mm}{b-a} \tag{3}$$

2.3.1.2 MLP-based model. The MLP network consisted of 201 input neurons (corresponding to the number of sampled points in the DOI ratio histograms), 2 fully-connected hidden layers with 256 and 128 neurons, respectively, and 3 output neurons corresponding to the slope ($k^{MLP}$) and intercept ($i^{MLP}$) for DOI calibration and the DOI resolution in millimetres ($DOI^{MLP}_{mm}$) (Fig. 4b). Hidden layers were activated with rectified linear unit (ReLU) functions and the mean squared error (MSE) between ground-truth data and predictions was used as the loss function. Network training was performed over 200 epochs with the Adam optimiser (learning rate = $1e^{-3}$). Training data were generated using the DOI calibration and resolution values ($k^{truth}$, $i^{truth}$, $DOI^{truth}_{mm}$) and DOI ratio histograms from Section 2.1.1 and Section 2.2.2, respectively. To mitigate bias in DOI resolution measurements, arising from increased scatter contributions in crystals farther from the source (Enríquez-Mier-y-Terán *et al* 2025), only the 230 crystals nearest to the source (~half the array) were used for training. To increase training data and to train the MLP with DOI ratio histograms from more oblique irradiation, data from 230 crystals from one detector in the panel were also included (Fig. 3; in purple). The DOI ratio histograms for these crystals were obtained from the uncollimated $^{22}$Na source irradiation detailed in Section 2.3.2. The MLP model was implemented using PyTorch v2.10 (Paszke *et al* 2019).

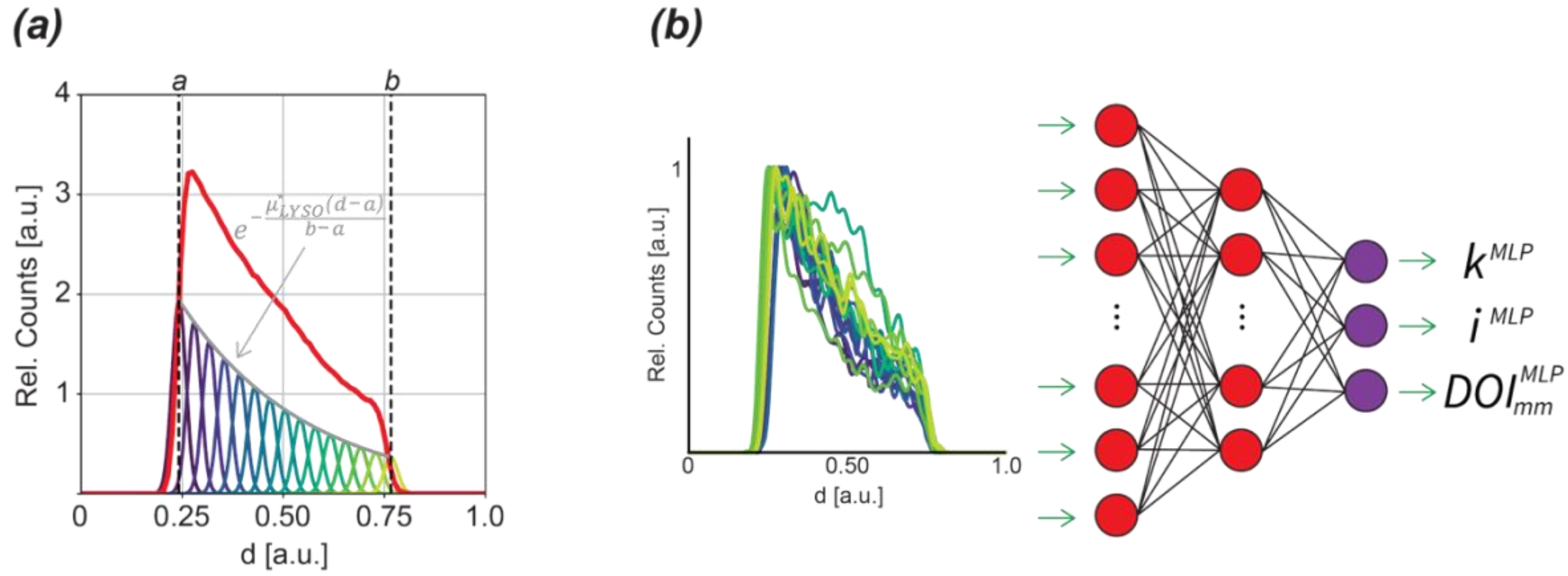

Fig. 4. (a) The DOI ratio histogram arising from uncollimated irradiation (red curve) is modelled as the convolution of the attenuation of the 511-keV photons in the crystal (grey line) and the intrinsic DOI response function of the detector, assumed to be Gaussian distributed. (b) The MLP is fed with crystal-wise DOI ratio histograms to extract the DOI calibration parameters ($k^{MLP}, i^{MLP}$) and the DOI resolution in millimetres ($DOI_{mm}^{MLP}$).

### 2.3.2. Detector Panel Calibration

To calibrate the detector panel, a one hour uncollimated irradiation was performed using a $^{22}$Na point source, with the source centred with respect to the detector panel front face and positioned 60 mm from it in the $z$ direction (Fig. 3). As in Section 2.1.2, LYSO intrinsic radiation data were collected for 1 hour for background subtraction of the energy spectra. For each of the 16 detectors crystal-wise DOI calibration and DOI resolution (in mm) were obtained using the MLP architecture (Section 2.3.1.2). Using the DOI calibration parameters, the data processing steps described in Section 2.1.2 were followed: crystal segmentation along the $z$ direction into 8 virtual crystals and extraction of depth-dependent 511-keV photopeak locations using fitted Gaussian distributions. Next, the $^{22}$Na point source was replaced by a $^{137}$Cs point source and the panel re-irradiated for an hour. Similarly, depth-dependent 662-keV photopeak locations were extracted crystal-wise. Per virtual crystal, the two photopeak locations (in ASIC units) were used to fit a detector saturation model (Schug *et al* 2019) to correct for the detector non-linear response as a function of energy. Lastly, energy resolution (after saturation correction) was measured crystal-wise, with the $^{22}$Na energy spectra in keV units.

## 3. Results

### 3.1. Single Detector Irradiation

Figure 5 shows the photopeak location and energy resolution as a function of detector depth for both collimated (slab-based) and uncollimated irradiation. For the uncollimated case, the box plots are centred on the virtual crystal bins.

Figure 5a shows the distribution of photopeak locations across the crystal array as a function of depth. Using uncollimated irradiation, the median photopeak location ranged from 140 ASIC units (arbitrary units) near the detector centre (10 mm) to 161 ASIC units near the detector ends, while for collimated irradiation it ranged from 141 to 159 ASIC units. Figure 5b shows the crystal-wise relative RMSE (RRMSE) between quadratic fits to the photopeak location as a function of depth for both irradiation methods. The mean and median RRMSE values across all crystals were 2.9 % and 1.0 %, respectively.

Energy resolution without saturation correction as a function of detector depth is shown in Figure 5c. The median energy resolution for uncollimated irradiation ranged from 10.9% to 11.8%, whereas for collimated irradiation it ranged from 9.6% to 11.6%. The interquartile range (IQR), aggregated across depths, was 4.2% and 3.2% for the uncollimated and collimated irradiation, respectively.

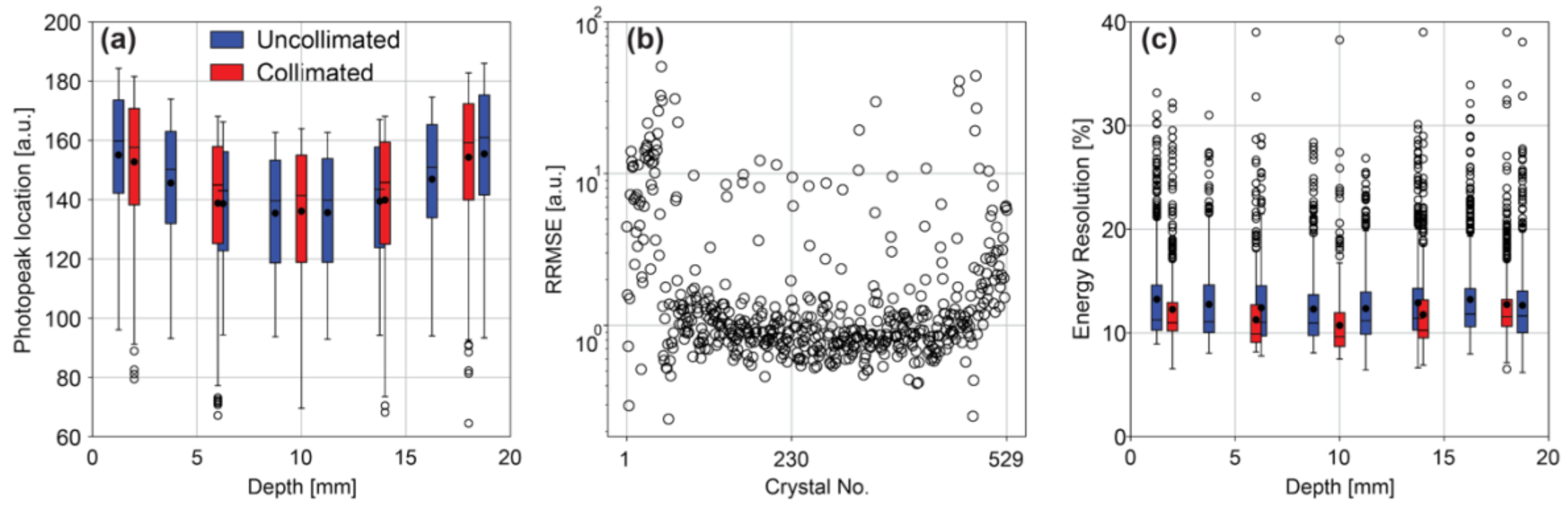

Fig. 5. (a) Photopeak location (in ASIC units) as a function of the detector depth for the uncollimated (blue) and collimated (red) irradiation methods. Box locations correspond to the depth centre of the virtual crystal. (b) RRMSE between the quadratic fits to the crystal-wise photopeak locations measured in each irradiation method. (c) Energy resolution (in %) for both irradiation methods.

## 3.2. Detector Panel Irradiation

### 3.2.1. DOI Calibration and Resolution

Figure 6 (top panel) shows the DOI calibration for the central crystal of detectors 01 (left), 02 (centre) and 03 (right) obtained using ground-truth collimated irradiation and uncollimated irradiation using the DOI model and MLP-based approach. Figure 6 (bottom panel) compares the RMSE performance of the DOI model and MLP-based approach with respect to ground truth. Table 1 shows the median RMSE across all crystals for both methods and each detector. For detector 03, 230 crystals were used for MLP training; these were excluded from the median RMSE calculation.

The DOI resolution (in mm) for each crystal in the three detectors is shown in Figure 7. For ground-truth collimated irradiation, the DOI resolution corresponds to the mean value measured across all irradiation depths. Table 2 summarizes the DOI resolution for the three detectors, obtained using ground-truth collimated and uncollimated irradiation with both the DOI model and the MLP-based approach.

Percentage relative errors between the DOI resolution obtained from ground-truth collimated irradiation and that from the MLP-based approach are shown in Figure 8. DOI resolution values in the distal half of the array (relative to the source) measured using collimated irradiation were artificially degraded due to increased inter-crystal scatter (Enríquez-Mier-y-Terán *et al* 2025). To mitigate this effect, the median absolute percentage error was computed using only the first 230 crystals. For detector 03, this metric may be artificially improved because the evaluation data overlapped with part of the training set. The median absolute percentage error for detectors 01, 02, and 03 were 6.4%, 6.3%, and 5.2%, respectively.

The DOI resolution for all detectors in the panel, measured using the MLP-based approach, is shown in Figure 9. Both the median and mean DOI resolution were 2.0 mm for all detectors except detector 07 which had median and mean values of 1.9 mm. IQR values varied from 0.08 to 0.15 mm across the detector panel.

Supplementary Figure S1 shows the DOI calibration and resolution for a detector block from a second detector panel.

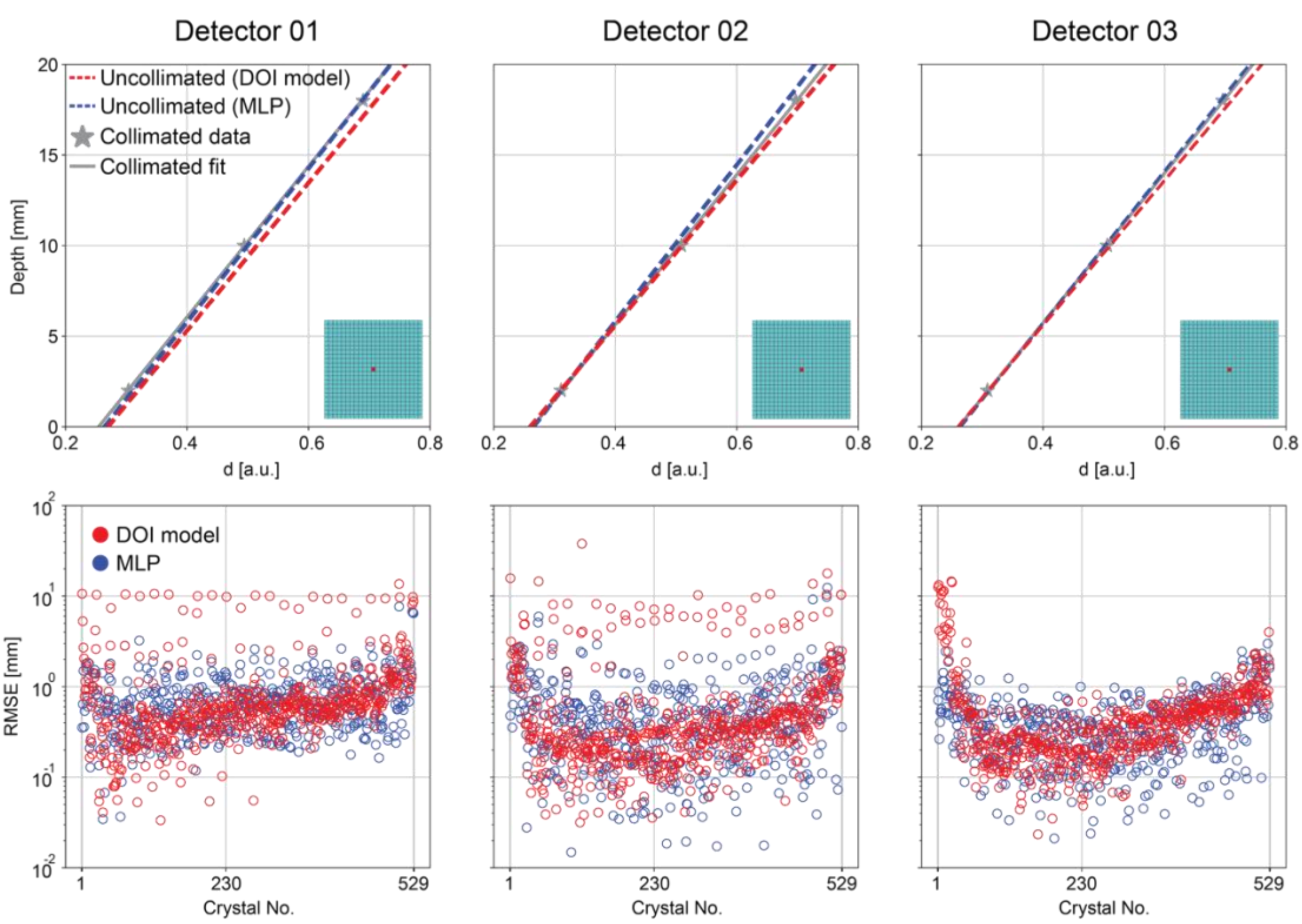

Fig 6. (top) DOI calibration for the central crystal in detector 01 (left), 02 (centre) and 03 (right) using ground-truth collimated irradiation at known depths (grey), and uncollimated irradiation using the DOI model (red) and MLP model (blue). (bottom) RMSE between the ground-truth DOI calibration and DOI model (red) or MLP model (blue).

Table 1. Median RMSE between ground-truth collimated DOI calibration and calibration using the DOI model and MLP model for three detectors in panel.

| Detector No. | DOI Model [mm] | MLP [mm] |
|---|---|---|
| 01 | 0.58 | 0.61 |
| 02 | 0.32 | 0.40 |
| 03* | 0.38 | 0.36 |

*Median RMSE for detector 03 excludes crystals using during MLP training

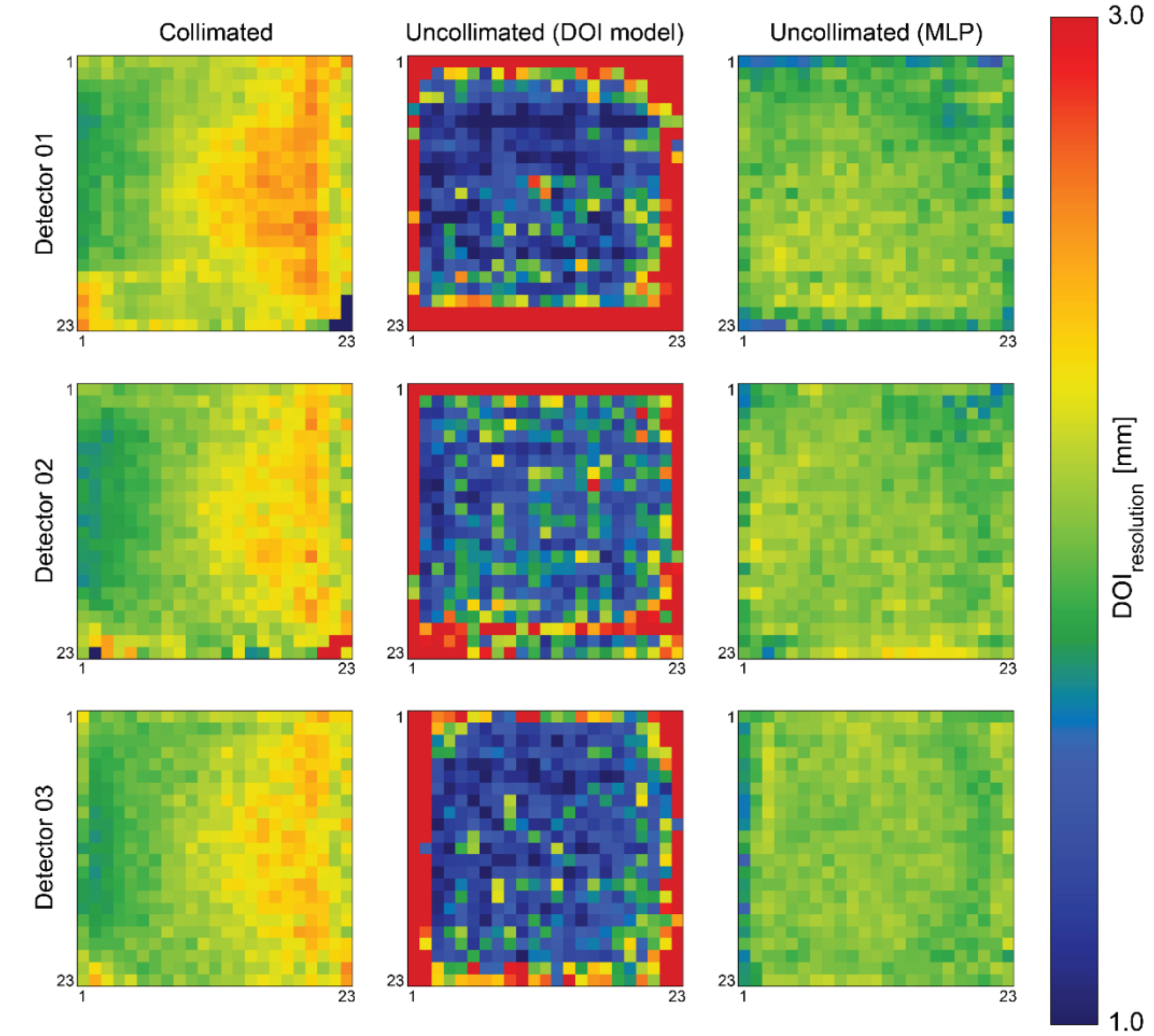

Fig. 7. Crystal-wise DOI resolution for detector 01 (top), 02 (middle) and 03 (bottom) measured using the ground-truth collimated irradiation approach (left) and DOI model (centre) and MLP-based (right) uncollimated irradiation approaches.

Table 2. Mean DOI resolution (± standard deviation) for detectors 01, 02 and 03 measured using the ground-truth collimated irradiation approach, the DOI model and the MLP-based uncollimated irradiation approaches.

| Detector | Collimated (Ground truth) [mm] | DOI Model [mm] | MLP [mm] |
|---|---|---|---|
| 01 | 2.2 ± 0.2 | 1.5 ± 0.2 | 2.0 ± 0.1 |
| 02 | 2.1 ± 0.2 | 1.7 ± 0.2 | 2.0 ± 0.1 |
| 03 | 2.1 ± 0.2 | 1.4 ± 0.1 | 2.0 ± 0.1 |

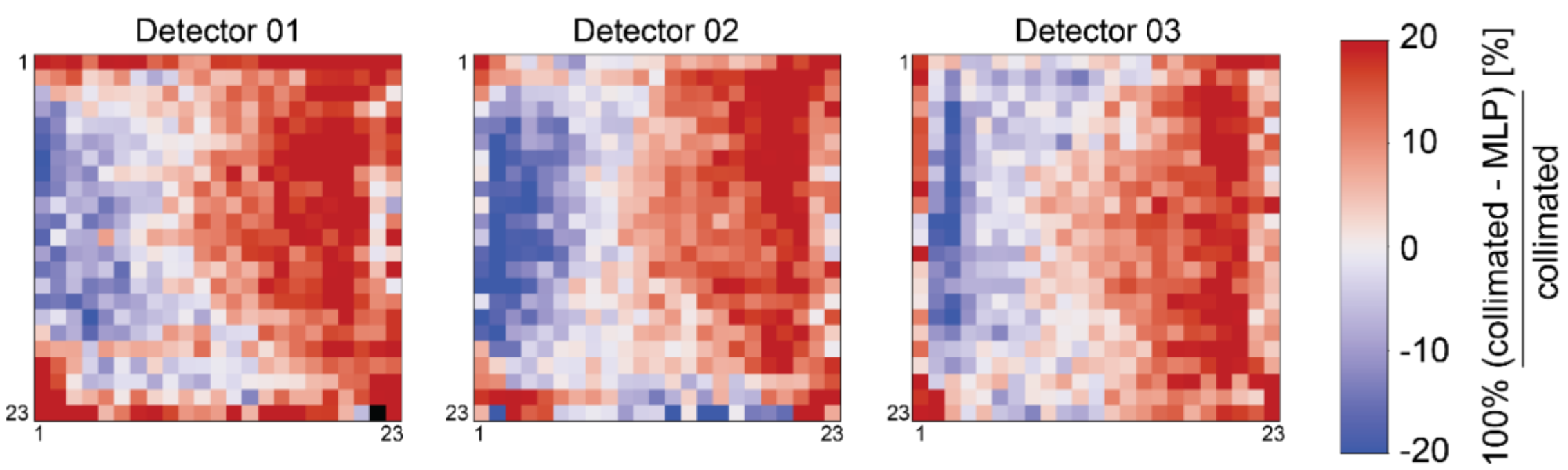

Fig. 8. Relative percentage error in DOI resolution from the MLP-based approach compared to ground-truth collimated irradiation for detector 01 (left), 02 (centre) and 03 (right).

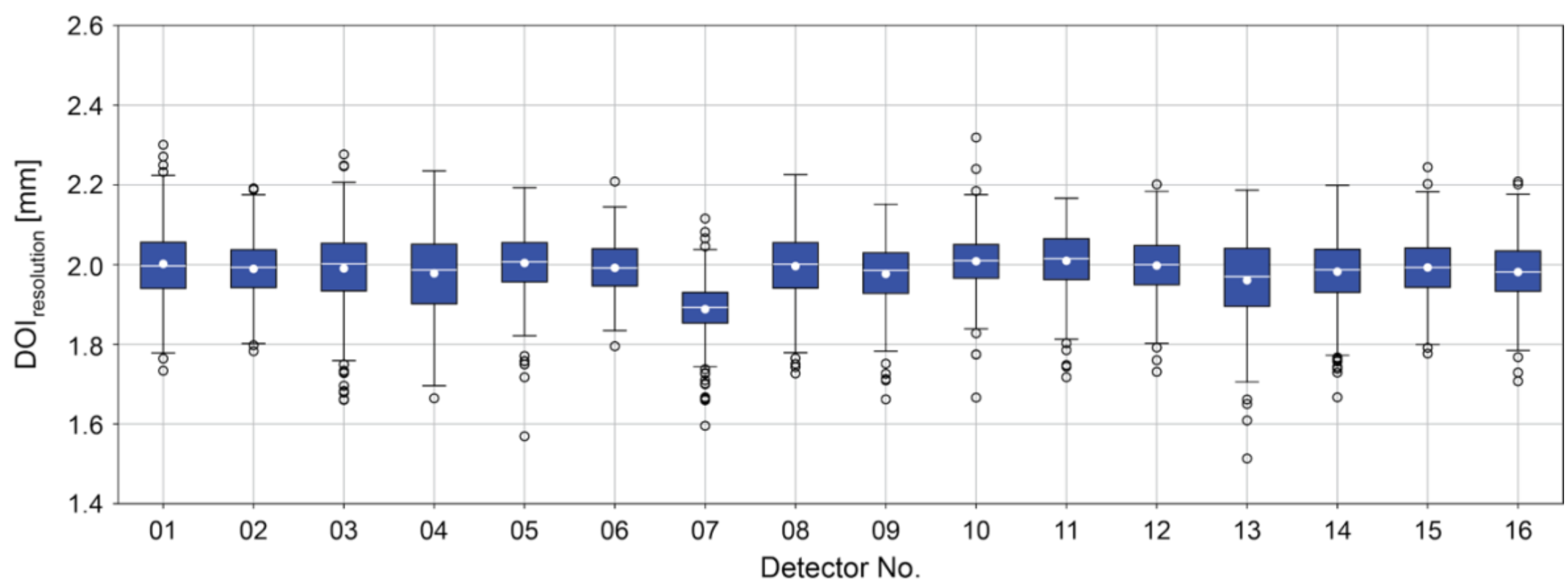

Fig. 9. Detector panel DOI resolution obtained using uncollimated irradiation and the MLP-based approach.

### 3.2.2. Energy Resolution

Figure 10 shows the energy resolution of the detector panel after saturation correction. For each crystal, the value displayed corresponds to the mean energy resolution measured across the eight virtual crystals. Table 3 reports the mean energy resolution of each of the 16 detectors. The detector panel mean energy resolution was 15.6 ± 4.0 %.

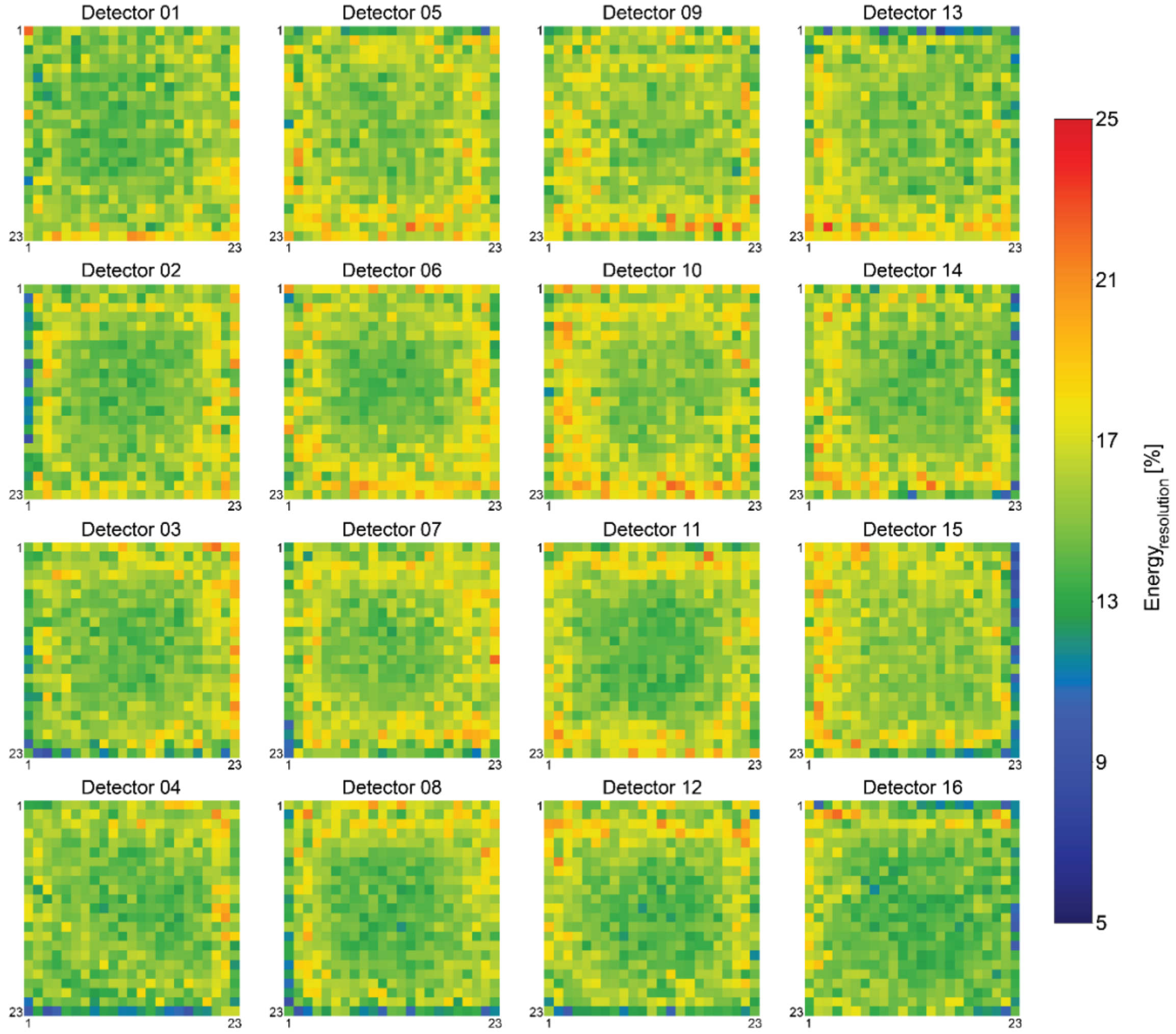

Fig 10. Mean energy resolution for every crystal in the detector panel.

Table 3. Mean detector energy resolution (± standard deviation).

| Detector | Energy Resolution [%] |
|---|---|
| 01 | 15.4 ± 3.9 |
| 02 | 15.5 ± 3.9 |
| 03 | 15.5 ± 3.9 |
| 04 | 15.1 ± 3.9 |
| 05 | 16.1 ± 4.0 |
| 06 | 16.0 ± 4.0 |
| 07 | 15.9 ± 4.0 |
| 08 | 15.3 ± 3.9 |
| 09 | 16.1 ± 4.0 |
| 10 | 16.4 ± 3.9 |
| 11 | 15.7 ± 3.8 |
| 12 | 15.6 ± 3.9 |
| 13 | 15.7 ± 4.3 |
| 14 | 15.6 ± 4.1 |
| 15 | 15.8 ± 4.1 |
| 16 | 14.7± 3.7 |

### 3.2.3. Detector Panel Flood Maps

Figure 11 shows the reconstructed detector flood maps from the $^{22}$Na source irradiation after selecting only events inside an energy window of 450 keV to 650 keV. Throughout the detector panel, some outermost crystals are lost due to low light collection, which in turn translates to noisy data and, ultimately, inaccurate calibrations.

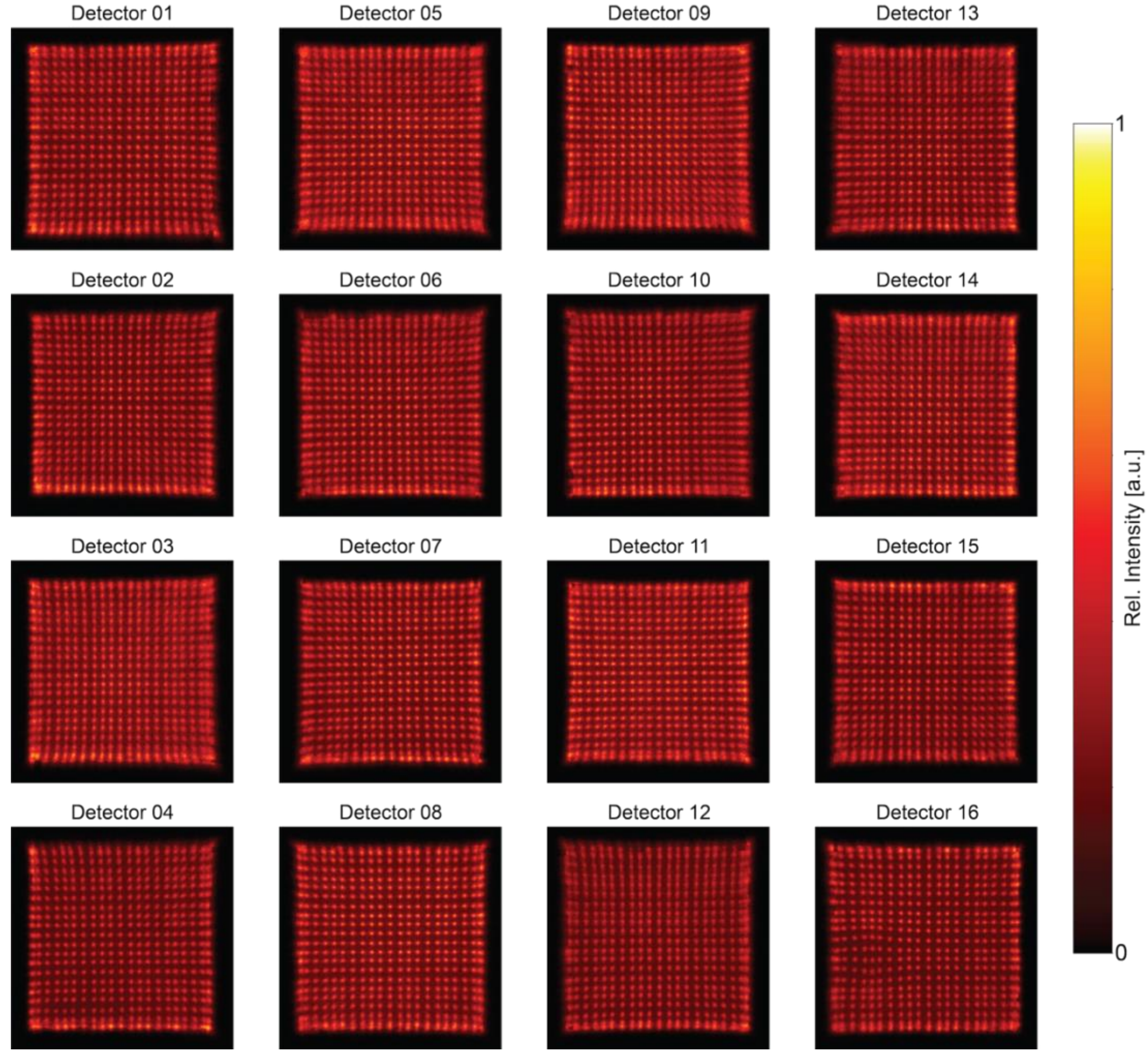

Fig. 11. Detector panel flood maps using events inside an energy window of 450-650 keV. Each flood map is normalised to its maximum pixel value.

## 4. Discussion

PET detectors used in sub-millimetre imaging applications must reliably estimate gamma-ray DOI to achieve accurate 3D positioning and depth-dependent energy calibration. However, traditional DOI calibrations have geometric constraints and/or inefficiencies that make them impractical for system-level deployment. In this work, we developed and validated a practical and robust approach for system-level DOI calibration for accurate gamma-ray 3D positioning and energy resolution evaluation.

### 4.1. Single Detector Experiments

Uncollimated front-face irradiation yielded photopeak locations as a function of the crystal depth comparable to those obtained using slab-based collimated irradiation. Because the virtual crystal bins from the uncollimated versus collimated setups had different width (2.5 mm versus 0.5 mm FWHM from the slab thickness), it was not feasible to perform a direct statistical comparison. However, comparison of quadratics fitted to the photopeak locations indicated a median RRMSE of 1%, suggesting that uncollimated irradiation provides a very reliable estimate of the depth dependence of the photopeak location.

The energy resolution (without saturation correction) measured using uncollimated irradiation was slightly degraded when compared to the slab-based collimated irradiation measurements. This is expected since the energy spectrum within a virtual crystal bin can be interpreted as the convolution of multiple spectra obtained using slab-based irradiation, each characterised by a narrower photopeak centred at slightly different ASIC units (i.e., corresponding to different depths). Although this effect could be mitigated by reducing the virtual crystal width, it would come at the expense of increased statistical noise in the energy spectra, particularly the last virtual crystal bin. In this work, eight virtual crystal bins were used as a trade-off between calibration accuracy and statistical noise but this could, in principle, be adjusted given a corresponding adjustment in acquisition time.

### 4.2. DOI Calibration and Resolution

For DOI calibration, median RMSE values comparing the physics-informed (Lerche *et al* 2008) and MLP-based approaches with the ground-truth collimated irradiation method (Table 1) slightly favour the physics-informed model. However, a Wilcoxon signed-rank test showed no statistically significant difference between the two approaches ($p \geq 0.1$) across the three detectors. For DOI resolution estimation, we note that the physics-informed model underestimates the DOI resolution for inner crystal rings and, conversely, overestimates the DOI resolution for outermost crystal rings. While the method has been widely used for DOI calibration in monolithic detectors, to our knowledge it has not been applied to pixelated detectors. In principle, the model should also be valid for pixelated detectors, as it estimates the DOI resolution (in mm) by scaling $\sigma_m$ according to Equation 3. This is equivalent to the approach used with collimated irradiation, where the standard deviation of the DOI-ratio distribution at a given crystal depth is converted to DOI resolution using the slope of the DOI calibration curve. Thus, the discrepancies measured here require further investigation—for example, by investigating how $\sigma_m$ changes as a function of the inter-crystal reflector, crystal cross-sectional area and thickness, and whether a correction factor should be included when scaling it to DOI resolution.

In contrast, the MLP-based approach outperforms the physics-informed DOI model in predicting DOI resolution. When considering only the first half of the crystal array (relative to the source location), the median absolute percentage error in the DOI resolution estimates for the three detectors remains in good agreement (≤10%) with gold-standard collimated irradiation. In the second half of the array, increased inter-crystal scattering degrades the DOI resolution and introduces a systematic bias, making full-array comparisons challenging. Although the MLP could be trained using data from the entire detector array, this approach would conflict with the primary goal of producing accurate, unbiased DOI resolution estimates. Future work will focus on improving the training strategy and/or implementing gamma-ray positioning algorithms capable of correcting inter-crystal scattering (Enríquez-Mier-y-Terán *et al* 2024). Even when using 230 crystals, some bias due to inter-crystal scatter may persist, while reducing the number of crystals per detector would limit the availability of labelled training data.

The comparable DOI calibration performance of the two approaches is noteworthy given the simplifying assumptions underlying the physics-informed model (Lerche *et al* 2008).The physics-informed model assumes that the DOI ratio histogram arises from the convolution of gamma-ray attenuation in the LYSO crystal with a Gaussian-distributed, depth-independent DOI resolution. However, these assumptions may not strictly hold since the gamma-ray attenuation model (Beer–Lambert law) assumes a parallel-beam geometry. While this may be valid for a small subset of crystals, most are subject to a significant contribution from gamma rays entering at oblique angles. In addition, we have previously shown that the DOI resolution is not independent of depth for this particular detector (Enríquez-Mier-y-Terán *et al* 2025). Nevertheless, these variations are relatively small and the assumption of a constant DOI resolution remains a reasonable approximation. This assumption is also implicitly adopted in the MLP-based approach.

Beyond its DOI resolution performance, the MLP-based approach offers practical advantages for detector calibration. In contrast to other machine-learning methods that estimate DOI on an event-by-event basis and typically require large training datasets, the proposed MLP model was trained using irradiation data from two sets of 230 crystals. This relatively simple training strategy may therefore assist the practicality of such models for DOI calibration. Furthermore, although the MLP model implemented here outputs parameters for a linear DOI calibration, it is readily extended to learn more complex, non-linear relationships between the detector metric and DOI.

### 4.3. Energy Resolution and Detector Flood Maps

Differences in crystal energy resolution were observed within each detector block, particularly in the central detector panel grid (i.e., detectors 06, 07, 10, and 11), where the outermost crystal rings exhibited slightly poorer energy resolution than the inner crystals (Fig. 10). Nevertheless, when the $23 \times 23$ crystal grid of each detector is divided into an inner grid ($17 \times 17$ crystals) and an outer grid (240 crystals), the mean energy resolution of the outer grid does not exceed that of the inner grid by more than 2% (data not shown). Assuming minimal inter-panel variability, the mean detector panel energy resolution measured here (15.6%) is comparable to that reported for other brain-dedicated small-animal PET systems (España *et al* 2014, Yamamoto *et al* 2016, Yang *et al* 2016, Kang *et al* 2023).

The detector panel assembled in this work is part of a broader effort to develop a motion-adaptive PET scanner for brain imaging of awake and freely moving small animals (Kyme *et al* 2017, Enríquez-Mier-y-Terán *et al* 2021). The need for mechanical stability during motion motivated the use of a fully glued panel and a single, continuous light guide rather than individual light guides for each detector block. However, this continuous design introduces inter-detector optical crosstalk, which negatively affects detector performance, including energy resolution. To mitigate this effect, optical barriers were created by machining sub-millimetre slits into the light guide and filling them with highly reflective white paint. The slit characteristics were not optimised through Monte Carlo simulations but were instead constrained by mechanical workshop capabilities. This represents a limitation of the present work, as a more informed light guide design could have been achieved using an optical Monte Carlo model of the detector panel which would, in turn, allow for a more detailed characterisation of inter-detector optical crosstalk.

For the analysis presented in this work, the flood maps were segmented into a 23 × 23 crystal grid using Voronoi diagrams. In some detectors, a small number of crystals were lost during calibration due to noisy data fits, which led to incorrect photopeak estimation. Because the crystal and SiPM arrays have similar footprint, the two outermost crystal rings do not share sufficient light with their corresponding SiPM arrays. Consequently, the positioning algorithm (Anger logic) cannot reliably distinguish these rings using only the eight detector signals. However, since light is shared between neighbouring detectors, future work will investigate positioning algorithms that incorporate signals from adjacent detectors, potentially enabling reliable discrimination of the outermost crystal rings. Visual inspection of the detector panel flood map in Figure 11 shows that, for most detector blocks, a grid of 21 × 21 crystals was fully resolved.

## 5. Conclusions

Accurate detector DOI calibration enables correct positioning of gamma rays along the crystal depth, and correction of other depth-dependent effects that degrade detector performance. However, the gold-standard DOI calibration is time-consuming and requires complex irradiation setups, making it impractical or impossible for calibrating high-resolution detector arrays at the system level. Using uncollimated irradiation, in which gamma rays are incident parallel or nearly parallel to the crystal depth direction, we demonstrate that a simple and easy-to-train MLP model can reliably estimate DOI calibration parameters and resolution with performance comparable to that of the gold-standard approach. Furthermore, we show that the same irradiation setup, combined with accurate DOI calibration, enables reliable estimation of the gamma-ray photopeak location in detectors where depth-dependent variation would otherwise degrade energy resolution. Finally, through the calibration of a 4 × 4 detector panel for ultra-high-resolution PET imaging, we demonstrate that this methodology is practical for system-level calibration of detector arrays. For the detector panel investigated, average energy and DOI resolutions of 15.6% and 2.0 mm, respectively, were achieved.

## Acknowledgements

The authors would like to thank Dr Julien Bec (University of California Davis, USA) for valuable feedback on the assembly of the detector panel.